# Ontology-based collaborative framework for disaster recovery scenarios

Sakkaravarthi Ramanathan[1], Aymen Kamoun[1,2], Christophe Chassot[1,2]
{sakkaravarthi, akamoun, chassot} @laas.fr
[1] CNRS ; LAAS ; 7 avenue du Colonel Roche, F-31077 Toulouse, France
[2] Université de Toulouse ; UPS, INSA , INP, ISAE ; LAAS ; F-31077 Toulouse, France

*Abstract*-This paper aims at designing of adaptive framework for supporting collaborative work of different actors in public safety and disaster recovery missions. In such scenarios, firemen and robots interact to each other to reach a common goal; firemen team is equipped with smart devices and robots team is supplied with communication technologies, and should carry on specific tasks. Here, reliable connection is mandatory to ensure the interaction between actors. But wireless access network and communication resources are vulnerable in the event of a sudden unexpected change in the environment. Also, the continuous change in the mission requirements such as inclusion/exclusion of new actor, changing the actor's priority and the limitations of smart devices need to be monitored. To perform dynamically in such case, the presented framework is based on a generic multi-level modeling approach that ensures adaptation handled by semantic modeling. Automated self-configuration is driven by rule-based reconfiguration policies through ontology.

## I. INTRODUCTION

Disaster is a natural hazard containing forest fire, earthquake, flooding, etc. It can be defined as a serious disruption of the functioning of a community or a society causing widespread human, material, economic or environmental losses which exceed the ability of the affected community or society to cope using its own resources. The total systematic coordination activities for the prevention and respectively the coverage of natural and man-made disasters are termed as disaster management activities [1]. Here, different actors like firemen, robots along with human volunteers operate together to save and rescue people. At the intervention area, wireless communication among the actors is more appropriate as it is not reliable to depend on the wired network. But the availability of this medium depends heavily on the actor's movements to save the victim. Also, the smart devices greatly vary in terms of e.g. processing and storage capabilities, energy consumption, and networking technologies. As communication resources could be lost or in demand, we need to have adaptive solutions to cope up the challenges posed by this environment. Thus, this work presents a design of ontology based collaborative autonomous system that provides adaptive solutions for achieving the mission tasks. Software agent is deployed at actor's device that triggers the control center if there is a problem. Once alarmed at control center, semantic model (ontology) and rules reconfigure dynamically its topology to ensure minimum guarantees to maintain the mission tasks.

This paper is structured as follows. Section 2 presents related work and section 3 outlines our disaster recovery scenario with brief introductions. Section 4 details our proposed approach. Section 5 focuses on our implementation in an adaptive architecture called FACUS. Section 6 concludes this work.

## II. RELATED WORK

Several researches have been carried out for collaborative systems and session management. A majority of these solutions deal with different aspects of collaboration. However, very few works treat specifically the problem of providing tools for building context-aware collaborative applications with dynamic reconfiguration of components at runtime.

We have studied existing synchronous collaborative systems such as TANGO [2], HABANERO [3], DOE200 [4] and DISCIPLE [5]. The main lack in these systems is that the all member roles cannot be changed dynamically during the collaborative activities. Consequently, these systems cannot support dynamic reconfigurations to maintain collaboration within structured sessions. Thus, model-based approaches are required in order to ensure the flexibility in the described systems. Other sessions models describe only three components of a session: users, tools and data flows [6, 7]. These proposed models establish collaborative sessions by monitoring members' activities inside groups. Some of them support dynamic change and provide a representation of the sessions but this representation is too specific to the model, what restricts its use in other collaborative systems.

Other ontologies-based works are proposed and applied to different problems of CSCW. Andonoff et al. [8] proposes ontology of high level protocols for agents' conversations. Ontologies are used in order to provide semantic to these protocols and to ensure automation of coordination in distributed systems. Garrido et al. [9] propose an MDA-based approach for modeling enterprise organization and developing groupware applications. The domain model is formalized through domain ontology in order to describe relations between actors sharing knowledge. Tomingas and Luts propose a semantic interoperability framework for data management like web services descriptions and ontologies [10].

The main disadvantage of classical collaborative systems is the lack of rigid deployment services of components or application that they offer. Components and applications are often deployed manually on the different machines and fixed at design time by a static way. This method cannot be applied to situations that need a high degree of adaptation, and in which even not known components should be deployed in advance.

Ontology has received great attention in the recent years, due to their use for knowledge representation in the Semantic Web domain. The Semantic Web was proposed by Tim Berners-Lee [11] in order to enrich data contained in the World Wide Web. The main idea is to add metadata in order to describe Web data (which is only human readable) in order to make it understandable by machines, thus enabling the automation of distributed processing over the Web. Metadata describing the semantics of contents is expressed in several languages such as RDFS (Resource Description Framework Schema, based on RDF) and OWL (Web Ontology Language). OWL is the Semantic Web standard for representing ontologies [12], which are common vocabularies allowing to model and represent knowledge. The main elements of ontologies are concepts, relations (between two concepts), individuals and axioms. All these elements are based on well-known formalisms such as Description Logics [13] in the case of OWL. Thus, knowledge can be automatically deduced by inference engines or (for example, Pellet [14]). These software elements can process ontology in order to make explicit the implicit knowledge that they use. Also, rules (expressed in SWRL, the Semantic Web Rule Language [15]) may be included in ontologies and processed by reasoners. Rules add some expressivity to OWL constructs.

We have chosen an ontology-based model because it constitutes a standard knowledge representation system, allowing reasoning and inference. Moreover, ontologies facilitate knowledge reuse and sharing through formal and real world semantics. Therefore, ontologies are high-level representations of business concepts and relations. These representations are close to developers' minds and therefore well suited to depict application level models. We have chosen to describe these models in OWL, the Semantic Web standard for metadata and ontologies.

For these reasons, ontologies seem a good choice for the representation of shareable collaboration knowledge. Standard tools are available and can be used for building and querying ontologies' instances. It also enables the sharing of collaboration concepts between several applications.

Moreover, the use of reasoning and rules such as SWRL is very useful. For example, they allow deducing, at run-time, the collaboration schema that corresponds to a given collaboration configuration required by the application.

Motivated by the above discussion, we present a novel ontology-based support for reconfigurable adaptive group communication architecture at control center. This approach improves the decision making that readily acts in a time-constraint situation.

## III. SCENARIO

The three major roles assigned to this scenario are: mission supervisor, coordinators, and field investigators. Each actor plays his own role and is associated with an identifier and the devices he uses. The performed functions are as follows:

- The supervisor's function is to monitor, manage and authorize actions to coordinators and investigators. This entity supervises the whole mission.
- Coordinator's job is to report to the supervisor. He manages the group of investigators during the mission and assigns tasks to each one of them.
- The investigator's role is to explore the operational field, observe, analyze, and report about the situation. They also take care of helping, rescuing and repairing.

Interactions between these actors are achieved by coordination and cooperation flows. Coordination flows take place between investigators and their coordinator and between the coordinators and the supervisor. Cooperation flows occur between the investigators within the same group (A2A type: fireman2fireman, robot2robot, etc.) or between the investigators of different groups (A2B type: robot2fireman, AAV2fireman, etc.). In case of A2A, a distinction is made between the flows such as cooperation notifications, cooperation requests and cooperation suggestions. In the case of A2B cooperation flows, the flows are: cooperation notifications and cooperation requests.

Figure 1 represents a supervisor, two coordinators, 3 firemen and 3 robots. The supervisor and coordinators have WiFi routers that are interconnected and the firemen and robots are connected to their respective coordinators through WiFi infra structure mode. If fireman 2 wants to communicate to robot1, the path will pass through coordinator, supervisor and coordinator of that robot. As the main goal is to help the victim, there exist four states. The first state is to trace the position of victim or own team members in case of danger. Once traced, the fireman moves to assist the injured person depending on its situation. This state sometimes involves no interference of coordinator but at times, the coordinator send request to fireman to locate the injured. Once the alarm is activated from the recovery team, the neighbour can recognize this situation and appropriate actions are taken. This state diagram helps the designer to understand the interactions between the coordinator and the investigator in an abstract way.

Once the connection is established between the actors, the system should be aware of the mission evolution. The system has to monitor, detect the changing environment and adapt to the evolution. It should handle and manage the connection between the control center as well as the other group members. If there is a failure in doing so, it should be smart enough to repair by activating suitable functions.

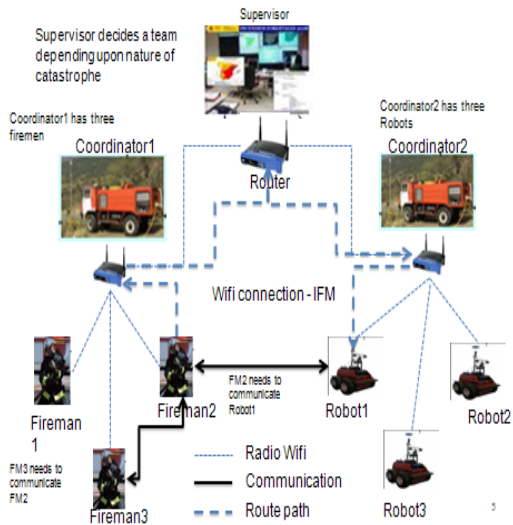

Figure 1: Public recovery application scenario

It is hard to diagnose problems manually in wireless communication system. A good solution is to automate a problem management task by continuously monitoring network condition, analyze the problem when it is detected, and taking adaptation actions for self-recovery. We have to detect, identify, isolate and determine the root cause of problems to recover the system. Problem detection measurements could be local and at the control center. Local measurements are resource usage monitoring such as CPU, memory and battery. It is used to estimate its own health status by investigator itself.

## IV. PROPOSED APPROACH

The adaptive techniques at two different states are explained in this section. We will elaborate the techniques used at the control center. At control center, the main feature is the clear partitioning of adaptive functionalities into different levels, in which each level only takes care of the functions that are most suitable to be concerned by it. Each level encapsulates issues into a specific model, thus abstracting complexity to a higher level. The modules at the higher level are abstract system representations that tend to resemble human activities, while lower ones are much closer to real implementations of abstractions supporting these activities. Relevant levels are identified and adaptation at the highest levels should be governed by changes in development of activity requirements. Adaptation at the lowest levels should be driven by execution context constraint changes.

In our work, we aim to support collaboration in distributed environments. New mechanisms are needed for managing session evolution and constraints changes. In our view, semantic web techniques are well suited to achieve this task. As far as we know, there is no existing collaborative system that use semantic for session management and dynamic component deployment. As thus, a semantic driven framework has been developed to enable session management and dynamic components deployment for collaborative systems.

The proposed framework is based on a multi-level modeling architecture (Figure 2). It ensures both high level and low level adaptation: The high level adaption depends on the configuration of different actors in the collaborative system such as their arrival, departure and role changes. While low level adaptation involve low level constraints such as energy level on the devices. SWRL rules are used in order to ensure the adaption at runtime. Some rules allow modeling the exchanged data flow between collaborating entities. Algorithms including different policies of low level constraints adaptation are also used in order to ensure the system robustness and the collaboration continuum.

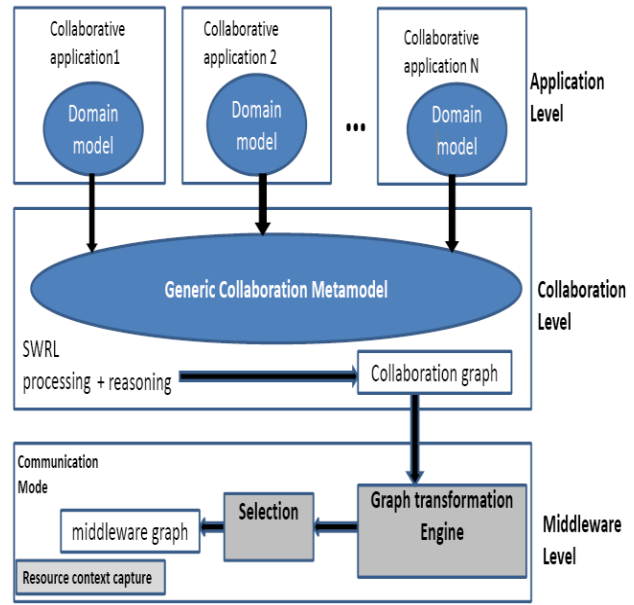

Figure 2: Multi-level architecture

### A. Application Level

This level represents the applications that need collaboration inside the group of devices. It includes software elements implementing the application's business, as well as user interfaces, etc. Among these elements, (at least) those relevant to collaboration are represented in the architectural model corresponding to this level of abstraction. Only collaboration-related elements of the application level model will be taken into account in the refinement process. Nevertheless, other business elements (non collaboration-related) are also included and can be used in order to represent the whole activity.

### B. Collaboration Level

This central level describes the way members in a group are organized within sessions, where they can send and receive data flows. The main issue is that of determining a high-level collaboration schema that meets the needs of application's collaboration. Hence, it supports collaborative sessions and can determine those elements needed to implement these sessions. In [9], an ontology model, containing generic collaboration knowledge as well as domain-specific knowledge, is proposed in order to enable architecture adaptation and to support spontaneous and implicit sessions inside groups of humans and devices.

The collaboration level model is a graph, inspired by dynamic collaboration ontology that models one or more session. A session is a set of data flows. Each data flow goes from a sender component to a receiver component (components are deployed on devices). Sender and receiver components may have text, audio or video as types. Flows are labelled with data types (audio, text and video) and the session to which they belong.

### C. Middleware Level

This level provides a middleware model that masks low-level details (like TCP sockets, UDP datagrams, IP addresses, multicast, etc.) in order to simplify the representation of communication channels. In actual fact, this level furnishes an abstract view of distributed systems, so that they become transparent for upper levels. For example, this model may be based on abstractions like Event-based Communications, Peer-to-Peer, Remote Procedure Calls or Remote Method Invocation.

Here, we have retained the Event-Based Communication (EBC) [16]. It represents a well established paradigm for interconnecting loosely coupled components and it provides one-to-many or many-to-many communication pattern. This model is a detailed graph containing a set of event producers (EP), event consumers (EC) and channel managers (CM) connected with push and pull links. Multiple producers and consumers may be associated through the same CM. Since this model represents a graph, it can also be expressed in the GraphML language.

### D. Application–Collaboration Refinement

Reasoning based on SWRL rules is used in order to implement the application-collaboration refinement. SWRL rules are applied to an instance of the domain ontology that extends the GCM. The proposed Generic Collaboration Meta-model includes a set of generic rules that express some relations and especially those which allow to infer a collaboration schema from the domain ontology instance. However, these rules are not sufficient for complete implementation of the refinement from the domain ontology to the collaboration ontology. Therefore the application designer has to specify additional rules in the domain application model which contains a part of the refinement process. The processing of the SWRL rules produces an instance of the Generic Collaboration ontology represented by a collaboration graph expressed in the OWL language. The application-collaboration refinement produces a single collaboration model from a given application level model.

### E. Collaboration – middleware Refinement

As the application level and the middleware level models are represented by graphs, graph grammar theories represent an appropriate formalism to handle the refinement process. We provide a graph grammar-based implementation of the refinement. We use a graph grammar, that addresses the refinement of a given activity level architecture to all possible EBC level architectures using Graph Matching Transformation Engine, GMTE [17]. The productions of this graph grammar consider data collaboration components (e.g. \texttt{ReceiverComponent} and SenderComponent) as non-terminal nodes and EBC entities (EPs, ECs and CMs) as terminal nodes. A session involving several senders and receivers is refined as a CM connected to several EPs and ECs.

### F. Middleware adaptation and selection

Given a set of possible middleware descriptor, it is necessary to select the best adapted architecture to be effectively deployed. We present here a procedure, Select() (Figure 3), that allows selecting one architecture depending on several parameters. This procedure uses the resources context (e.g. energy level, bandwith, etc.) to discard the set of architecture that cannot be deployed within the current resources levels. Among the set of selected architectures, the best configuration is selected by processing a set of defined policies.

```
1  Select(Policy)
2  {
3  Let A_{n,p} ∈ 𝔸_n, p ∈ ℕ
4  Let C denote the context attributes
5  Select S_1 = {A_{n-1,k} ∈ 𝔸^p_{n-1}, k ∈ ℕ such that:
       Context_Adaptation(A_{n-1,k}, C) ≥
       Context_Adaptation(X, C), ∀X ∈ 𝔸^p_{n-1}}
6  if card(S_1) ≠ 1
7     if Policy = Dispersion
8        Select S_2 = {A_{n-1,k} ∈ S_1, k ∈ ℕ such that:
          Dispersion(A_{n-1,k}) ≥ Dispersion(X), ∀X ∈ S_1}
9     if Policy = Distance
10       Let A_{n,p} and A_{n,q} ∈ 𝔸_n, p, q ∈ ℕ
11       Let A_{n-1,p} the current mapping
            at level n − 1 of A_{n,p}
12       Select S_2 = {A_{n-1,k} ∈ 𝔸^q_{n-1}, k ∈ ℕ such that:
13       Relative_Cost(A_{n-1,p}, A_{n-1,k}) ≤
            Relative_Cost(A_{n-1,p}, X), ∀X ∈ S_1}
14    if card(S_2) ≠ 1
15       Select any configuration from S_2
16 }
```

Figure 3: Selection procedure

The choice of a middleware descriptor must take into account the resources context at first. For that, The *context_Adaptation()* function (Figure 3, line 5) which is a generic function that manages the resources context is used. It can express the availability level of a given resource (bandwidth, memory, energy level, etc.). This function is used for two purposes: first, it allows discarding not adapted architectures that cannot be deployed within the current resources context. Second, it allows selecting the best adapted architectures to that context. This function assigns a value to a given architecture depending on its degree of adaptation to the current context. If the architecture is not compatible with the current context, its value will be −1. Otherwise, it will receive a positive value. Best suited architectures will be assigned higher values.

When several architectures have been assigned the same value of *Context_Adaptation()*, a policy (indicated by the parameter Policy) is used by the Select() procedure in order to select the optimal configuration. If the chosen policy is *Dispersion*, the selection is based on maximizing the function Dispersion() (Figure 3, line 8). This generic function is defined as the number of software components deployed per node. If the chosen policy is *Distance*, the selection minimizes the distance between two middleware descriptors. This is performed using the function Relative_Cost() (Figure 3, line 13).

### G. Reconfiguration Rules

To make this system adaptable, reconfiguration rules are needed to adapt the ontology instance to the current situation. As events play a major role in our application, the transformation of entities needs to be triggered. Here, events could occur at activity level, i.e, addition of new participants, changing an action, transfer of a participant from one group to another, new connection between investigators from different groups, etc. The events could also take the form of resource context changes, e.g, parameters like the energy of a device, bandwidth range, CPU processing capacity, RAM availability etc. We use SWRL rules to define our adaptation policy. The application designer defines these rules according to context changes he wants to handle. If there is no solution for an event at a particular level, then it triggers the higher level. In SWRL, the head points to the adaptation transformations whereas the body indicates the context of ontology elements. This reconfiguration rules are really useful for adapting the scenario dynamically.

## V. IMPLEMENTATION

We implemented our work using the proposed Framework that supports semantic adaptation enabling the awareness of the presence/absence, roles and tasks of collaborators. This framework is based on a generic multi-level modeling approach that ensures multi-level adaptation. A generic collaboration model, based on Semantic Web technologies is proposed in order to support real-time collaboration between groups of participants working together in different tasks. The framework defines common interfaces for collaborative systems to enable the management of cooperative actions.

In the implementation, we have chosen one group composed of a supervisor, two coordinators and two firemen investigators to show the adaptation. Initially, the supervisor and the two coordinators are connected using WiFi infrastructure. Also, the connection between the firemen coordinator and the two investigators (fireman1 and fireman2) are connected in WiFi infrastructure mode.

### A. Experiments

The Table1 represent the variation of the energy level of each device during the two phases: phase1 and phase2. We notice that the energy level of the investigator 1' device decreases to 50 during the phase 2.

| Role | Device (IP) | Energy level phase1 | Energy level phase2 |
|---|---|---|---|
| Supervisor | 10.193.255.1 | 86 | 86 |
| Fireman coordinator | 10.193.255.100 | 90 | 90 |
| Robot coordinator | 10.193.255.200 | 79 | 79 |
| Fireman 1 | 10.193.255.143 | **93** | **50** |
| Fireman 2 | 10.193.255.146 | 88 | 88 |

Table 1 : Devices' energy levels

*Phase 1*

This phase represents the initial state of the collaborating actors. The figure 4 represents the initial configuration: The team Team1 has 5 members: a supervisor having the IP address: 10.193.255.1, a fireman coordinator having the ip address 10.193.255.100, a robot coordiantorhaving the ip address 10.193.255.200 and two investigators having ip address: 10.193.255.143 and 10.193.255.146.

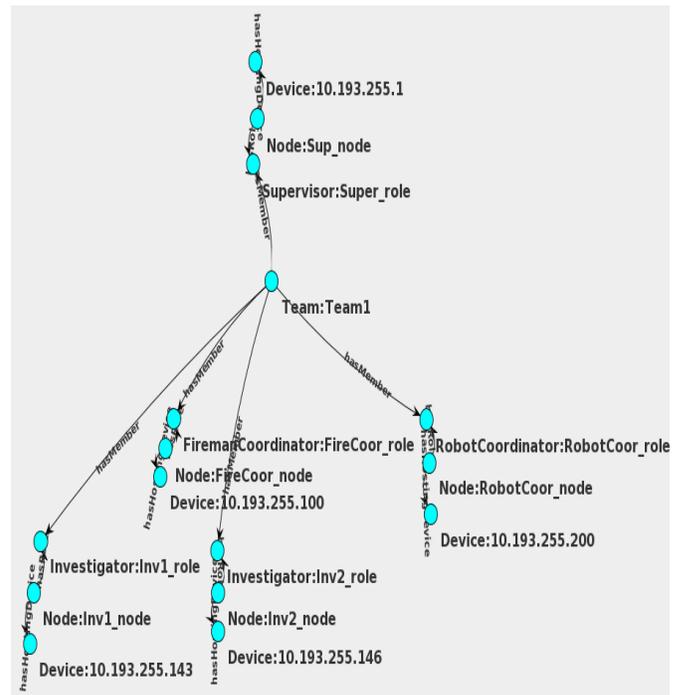

Figure 4: The application state

```
Investigator(?inv) ∧ sessions:Node(?ninv) ∧ sessions:Device(?dinv) ∧
sessions:hasRole(?ninv, ?inv) ∧ sessions:hasHostingDevice(?ninv, ?dinv)
∧ FiremanCoordinator(?coo) ∧ sessions:Node(?ncoo) ∧
sessions:hasRole(?ncoo, ?coo) ∧ sessions:Device(?dcoo) ∧
sessions:hasHostingDevice(?ncoo, ?dcoo) ∧
sessions:hasSameSSID(?dinv, ?dcoo) ∧ hasSignalWith(?dinv, ?dcoo) ∧
 sessions:belongsToSameGroup(?inv,?coo) ∧ differentFrom(?inv, ?coo) ∧
sessions:belongsToGroup(?coo, ?t) ∧ hasFiremanCordInvSession(?t, ?s)
∧ swrlx:createOWLThing(?af1, ?inv) ∧
swrlx:createOWLThing(?af2, ?inv) →
sessions:AudioFlow(?af1) ∧ sessions:hasSource(?af1, ?ncoo) ∧
sessions:hasDestination(?af1, ?ninv) ∧ sessions:belongsToSession(?af1,
?s) ∧ sessions:AudioFlow(?af2) ∧ sessions:hasSource(?af2, ?ninv) ∧
sessions:hasDestination(?af2, ?ncoo) ∧
sessions:belongsToSession(?af2, ?s)
```

Figure 5: an example of SWRL rule

The figure 6 represents the high level collaboration graph that models the actor needs. Given the configuration of the application level (figure 4), a set of SWRL rules allows inferring the collaboration schema that describe the necessary components (receiver and sender) for each node. The figure 5 describes the SWRL rule allowing establishing the communication between the fireman coordinator and investigators by creating senders and receivers.

This rule may be described as follows: if the application contains an investigator and a fireman coordinator, and they have the same SSID, and they can bee connected (*hasSignalWith*), and tey belong to the same group, so two flows can be created: the first one sent by the investigator and received by the fireman coordinator and the second one will be sent by the fireman coordinator and received by the investigator. Other SWRL rules are created in order to ensure the connection between the robot coordinator and the associated investigators and between the supervisor and the coordinators.

After having processing the SWRL rules, we obtain the collaboration graph represented by the figure 6. This graph represents exchanged data flows between senders and receivers. Each vertex is composed of 5 attributes: the vertex' id, the vertex' type (sender or receiver), the ip address, the exchanged data flow and the name of the session in which the node collaborate. For the created session between the supervisor (10.193.255.1) and the two coordinators (10.193.255.100 and 10.193.255.200), 2 receiver components allow the supervisor to receive data flow from the 2 coordinators and a sender component to send data flows to the two coordiantors. The same type of components will be provided for the created session between the fireman coordinator (10.193.255.100) and the two investigators (10.193.255.143 and 10.193.255.146).

Given the collaboration descriptor, the GMTE will generate all possible EBC-based middleware graphs deploying the ECs, EPs and the CMs on different devices. It will replace each receiver by an *EventConsumer* (EC) and each sender by an *EventProducer* (EP). A broker called *ChannelManager* (CM) will be created for each session. After that, the selection procedure selects the best adapted middleware graph depending on several parameters such as the energy levels.

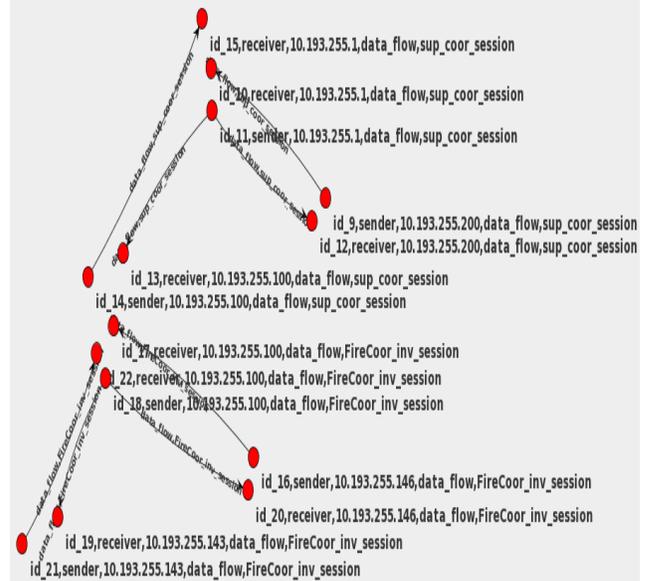

Figure 6: the collaboration graph

The figure 7 represents the selected middleware graph for the pahse1. Each vertex in this graph is composed of 6 attributes: the vertex' id, the component type (EP, EC or CM), the exchanged data flow, the session name and the ip address of the device on which the component will be deployed.

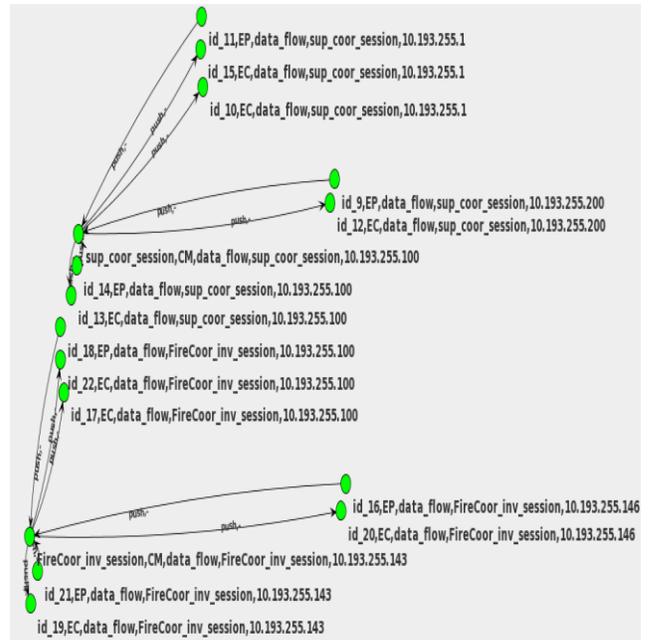

Figure 7: The selected middleware graph

As represented by this figure, EPs have to be deployed for each sender; ECs have to be deployed for each receiver and CMs have to be deployed for each session. For the session including the supervisor and the two coordinators (*sup_coor_session*), the CM will be deployed on the fireman coordinator' machine (10.193.255.100) because the energy levels are taken into account by the selection procedure (the fireman coordinator has a higher energy level (90) than the supervisor (86) and the robot coordinator (79) (table1)). For the session including the fireman coordinator and the two investigators (Fire*coor_inv_session*), the CM will be deployed on the Fireman2' machine (10.193.255.143) because the energy levels are taken into account by the selection procedure (the Fireman1 has a higher energy level (93) than the fireman coordinator (90) and the fireman2 (88) (table1)).

*Phase2 (adaptation)*

In phase 2, the energy level of the fireman 1 decreases from 93 to 50. The selection procedure has to select a new middleware graph more adapted to the current resources context. The figure 8 shows the new selected middleware graph. In this graph, the CM which was deployed on the fireman1'device (10.193.255.143) will be moved to the fireman2'device (10.193.255.146).

*Phase3 (arrival of new robot)*

The figure 9 represents the middleware graph that contains all required communication component after the arrival of the new robot (10.193.255.202). This investigator has the same SSID as the robot coordinator, so a new session is created enabling the communication between them. This middleware descriptor is obtained after running the adaptation process which generates the application descriptor, the collaboration descriptor and the middleware descriptor after the processing of the selection procedure. Here, we show only the middleware graph for simplicity.

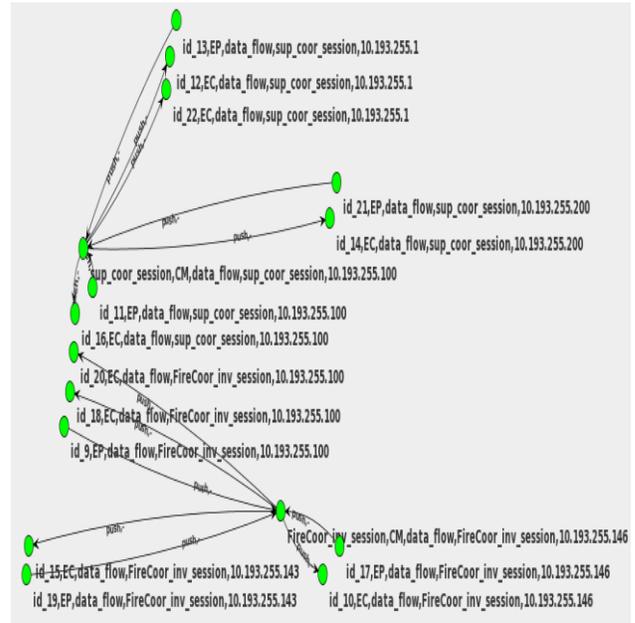

Figure 8: The middleware graph after adaptation

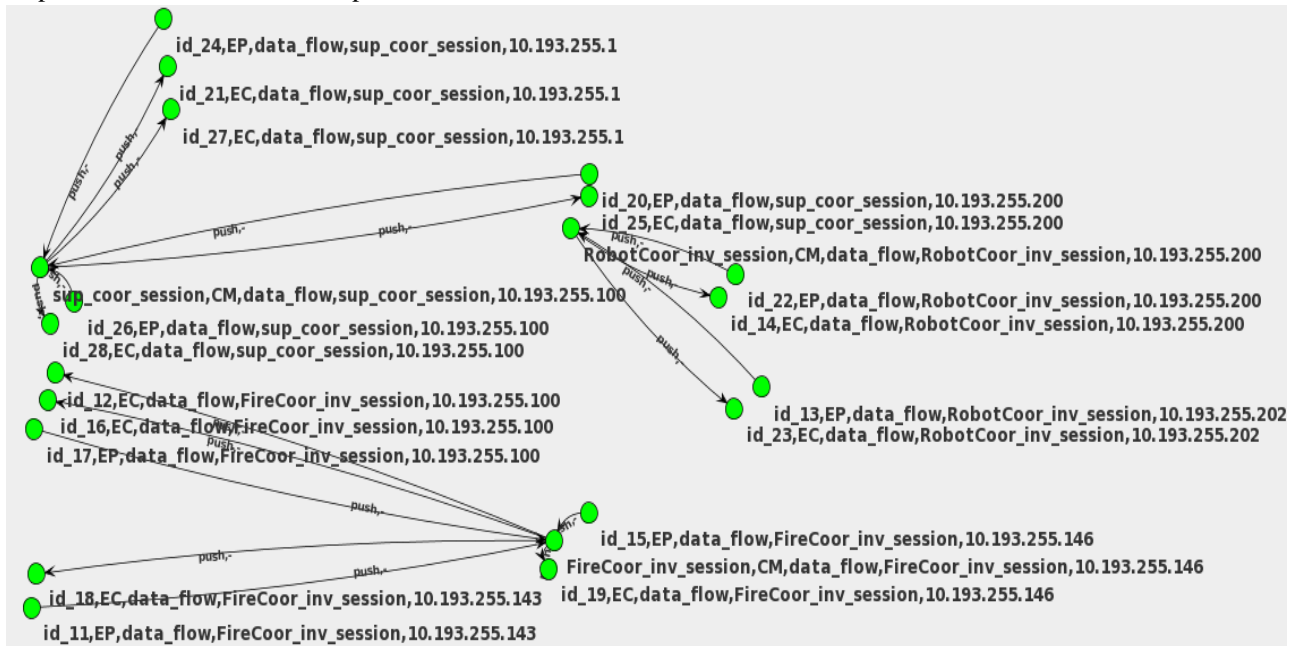

Figure 9: The middleware graph after arrival of new robot

## VI. CONCLUSION

In this work, a multi-level modeling approach designed to support collaborative adaptation for public safety and disaster applications has been detailed. Here, the whole adaptive process has been divided into different levels. Throughout the higher levels, ontology has been used where as event-based communication has been retained at the middleware. If a change arises in the environment, reconfiguration can be achieved by using SWRL rules at run-time to handle the mission evolving conditions.

Adaptation is achieved by auto-configuring the system and its associated components after detecting a change in the mission such as new arrival of a actor and problem caused due to resource constraint in communication device.

ACKNOWLEDGEMENT

This research is supported by the French project ROSACE (RObots et Systémes Auto-adaptatifs Communiquants Embarquès).